\documentclass{emulateapj}

\shorttitle{Photo-z with shape parameters}
\shortauthors{Singal et al.}

\slugcomment{Accepted to PASP}

\begin{document}

\title{THE EFFICACY OF GALAXY SHAPE PARAMETERS IN PHOTOMETRIC REDSHIFT ESTIMATION:\\ A NEURAL NETWORK APPROACH}

\author{J. Singal\altaffilmark{1}, M. Shmakova\altaffilmark{1}, B. Gerke\altaffilmark{1}, R.L. Griffith\altaffilmark{2}, J. Lotz\altaffilmark{3}}

\altaffiltext{1}{Kavli Institute for Particle Astrophysics and Cosmology\\SLAC National Accelerator Laboratory and Stanford University\\2575 Sand Hill Road, Menlo Park, CA 94025}
\altaffiltext{2}{Jet Propulsion Laboratory\\ MS 169-327, 4800 Oak Grove Dr.\\ Pasadena, CA 91109}
\altaffiltext{3}{Space Telescope Science Institute\\3700 San Martin Drive\\Baltimore, MD, 21218}

\email{jsingal@stanford.edu}

\begin{abstract}
We present a determination of the effects of including galaxy morphological parameters in photometric redshift estimation with an artificial neural network method.  Neural networks, which recognize patterns in the information content of data in an unbiased way, can be a useful estimator of the additional information contained in extra parameters, such as those describing morphology, if the input data are treated on an equal footing.  We use imaging and five band photometric magnitudes from the All-wavelength Extended Groth Strip International Survey.  It is shown that certain principal components of the morphology information are correlated with galaxy type.  However, we find that for the data used the inclusion of morphological information does not have a statistically significant benefit for photometric redshift estimation with the techniques employed here.  The inclusion of these parameters may result in a trade-off between extra information and additional noise, with the additional noise becoming more dominant as more parameters are added.

\end{abstract}

\section{Introduction} \label{intro}

Obtaining sufficiently accurate photometric redshift estimates is of the utmost importance for the current and coming era of large multi-band extragalactic surveys (see e.g. \citet{Huterer06} for a recent review).  Unlike spectroscopic redshift determination, photometric redshift estimation (photo-z) is highly subject to systematic errors and confusion because the spectral information of a galaxy is limited to the magnitude or flux in a number of wavelength bands.

Photo-z estimation techniques have traditionally been divided into two main classifications.  So-called ``Template fitting'' methods, for example the popular {\it Lephare} package as described in \citet{lephare} and \citet{Arnouts99}, and {\it Bayesian Photometric Redshift (BPZ)} as described in \citet{bpz}, involve correlating the observed band photometry with model galaxy spectra and redshift, and possibly other model properties.  In contrast, so-called ``Empirical'' or ``Training set'' methods, such as artificial neural networks \citep[e.g. {\it ANNz},][]{annz} and boosted decision trees \citep[e.g. {\it BDT},][]{Gerdes10}, develop a mapping from input parameters to redshift with a training set of data in which the actual redshifts are known, then apply the mappings to data for which the redshifts are to be estimated.  There are advantages and disadvantages to each class of methods.  Template fitting methods require assumptions about intrinsic galaxy spectra or their redshift evolution, and empirical methods require the training set to be `complete' in the sense that it is representative of the target evaluation population in bulk in all characteristics.  

In regard to photo-zs, science goals such as using weak lensing for cosmology are most affected by the number of outliers - those objects whose estimated photo-zs are far from the actual redshifts \citep[e.g.][]{Hearin10}.  In general, data sets with bands extending into the infrared (e.g. J, H, and K bands) have more accurate photo-z estimation and fewer outliers.  However, most upcoming large surveys, such as the Large Synoptic Survey Telescope \citep[LSST,][]{LSSTover}, will have optical and near-infrared data only.  

It is a reasonable hypothesis that galaxy morphology and redshift are correlated in such a way that the addition of morphological information could improve photo-z estimation.  Reasons include the larger frequency of mergers at higher redshifts, and, perhaps more importantly, the general evolutionary trend from spiral to elliptical shapes.  

The inclusion of morphological parameters in photo-z estimation has also been studied by \citet{Tag03} with an artificial neural network determination, and by \citet{VC2006} and \citet{WS2006} with other methods.  All three works use Sloan Digital Sky Survey (SDSS) data.  \citet{Tag03} find possible modest improvement with the inclusion of shape information, although they restrict their analysis to quite low redshift (z$\leq$0.7) galaxies.  \citet{WS2006} consider several empirical methods and show marginal improvement for some methods with the addition of morphological information.  \citet{VC2006} claim an improvement of between 1 and 3 percent in the RMS error in photo-z determination, however it is not noted whether this result is significant and the method of photo-z estimation is not discussed.  We note that SDSS galaxy photometric data is a bit unusual in the context of data that will be used to constrain cosmological parameters from surveys such as LSST, in that SDSS photometric data has a greater representation of nearby galaxies and thus fewer potential outliers.

In this work, we explore the efficacy of adding parameters describing the morphological information of galaxies, in the context of a neural network estimation technique for photo-zs.

\section{Data set and shape parameters} \label{dataset}

For this analysis we desire data with magnitudes in a number of optical bands, spectroscopic redshifts, and enough imaging resolution to determine morphological parameters.  We use observations of the Extended Groth Strip from the the All-wavelength Extended Groth Strip International Survey (AEGIS) data set \citep{EGS}, which contains photometric band magnitudes in u, g, r, i, and z bands from the Canada-France-Hawaii Telescope Legacy Survey \citep[CFHTLS,][]{Gwyn08}, imaging from the Advanced Camera for Surveys on the Hubble Space Telescope \citep[HST/ACS,][]{K07}, and spectroscopic redshifts from the DEEP 2 survey using the DEIMOS spectrograph on the Keck telescope.  The limiting i band AB magnitude of the CFHTLS survey is 26.5, while that of HST/ACS is 28.75 in V (F606W) band, and that of DEEP2 is 24.1 in R band.  

From the HST/ACS imaging data in two bands, V (F606W) and I (F814W), we form a set of parameters characterizing the morphological properties of the galaxies as follows:

1. The Concentration C: $C=5 \, log \, {{r_{80}} \over {r_{20}}}$
This parameter defines the central density of the light distribution with radii $r_{80}$ and $r_{20}$ correspondingly 80\% and 20\% of the total light.

2. The Asymmetry A:  $A = { {\Sigma_{x,y} \vert I_{(x,y)}-I_{180(x,y)} \vert} \over {2\Sigma_{x,y} \vert I_{x,y} \vert} } -B_{180} $
This parameter characterizes the rotational symmetry of the galaxy's light, with $I_{(x,y)}$ being the intensity at point (x,y) and $I_{180(x,y)}$ being the intensity at the point rotated 180 degrees about the center from (x,y), with $B_{180}$ being the average asymmetry of the background calculated in the same way.  It is the difference between object images rotated by 180$^{\circ}$.

3. The Smoothness S:  
$S = { {\Sigma_{x,y} \vert I_{(x,y)}-I_{S(x,y)} \vert} \over {2\Sigma_{x,y} \vert I_{x,y} \vert} } -B_{S} $.
The smoothness is used to quantify the presence of small-scale structure in the galaxy.  It is calculated by smoothing the image with a boxcar of a given width and then subtracting that from the original image.  In this case $I_{(x,y)}$ is the intensity at point (x,y) and $I_{S(x,y)}$ is the smoothed intensity at (x,y), while $B_S$ is the average smoothness of the background, calculated in the same way.  The residual is a measure of the clumpiness due to features such as compact star clusters.  In practice, the smoothing scale length is chosen to be a fraction of the Petrosian radius.

4. The Gini coefficient G: \\
$G={ {1} \over {\bar{X} \, n(n-1)} } \Sigma^n_i (2i-n-1) \, X_i$,
describes the uniformity of the light distribution, with $G=0$ corresponding to the uniform distribution and $G=1$ to the case when all flux is concentrated in to one pixel.  $G$ is calculated by ordering all pixels by increasing flux $X_i$. $\bar X$ is a mean flux and $n$ is the total number of pixels.

5. $M_{20}$: $M_{20}=log \Sigma M_{i}/M_{tot}$,
is the ratio of the second order moment of the brightest 20\% of the galaxy to the total second moment.  This parameter is sensitive to the presence of bright off-center clumps.

6.  The Ellipticity $\varepsilon$:  $\varepsilon = 1 \, - \,  {{b} \over {a}}$.
The values $a$ and $b$ are the semi-major axis and semi-minor axis of the galaxy.

A number of these parameters are discussed in e.g. \citet{Scar06}.  The remaining two parameters are two of the fitting parameters to the S\'ersic profile form $\Sigma(r) = \Sigma_e\,e^{-k \vert (r/r_e)^{(1/n)}-1 \vert}$ \citep[e.g.][]{GD05}, where $\Sigma_e$ is the surface brightness at radius $r_e$ and $k$ is defined such that half of the total flux is contained within $r_e$ :

7. The S\'ersic power law index $n$

and

8. $r_e$, the effective radius of the S\'ersic profile.

Morphological parameters C, A, S, G, and $M_{20}$ are determined for these galaxies in \citet{Lotz08}, while $\varepsilon$, n, and $r_e$ are determined by \citet{Griffith11} using the Galfit package \citep{Peng02,Hausler07}.  For this analysis we require a magnitude in each band, a spectroscopic redshift, and sufficient HST/ACS image resolution to construct all eight shape parameters.  A total of 2612 galaxies spanning redshifts from 0.01 to 1.57, with a mean redshift of 0.702 and a median of 0.725, and i band magnitudes ranging from 24.43 to 17.62, are in the data set used here.  The redshift distribution of this particular set of galaxies arises because of the intentional construction of the portion of DEEP2 spectroscopic catalog within the AEGIS survey to have roughly equal numbers of galaxies below and above z=0.7; therefore it is not an optimized training set for a generic photometric data evaluation set, although a more optimized training set for any given photometric data evaluation set could be constructed from it.  We emphasize that because in this analysis random subsets from the same 2612 galaxy catalog are used for training and evaluation, the representativeness of the training set is not an issue here for this analysis.

Template-based photometric redshifts estimations for all of the galaxies used in this analysis have been reported in \citet{lephare}.  This estimation also provides a most likely galaxy type among template spectra corresponding to elliptical, Sbc, Scd, Irregular, or Starburst.

\section{Morphological principal components}\label{pc}

\begin{figure}
\includegraphics[width=3.5in]{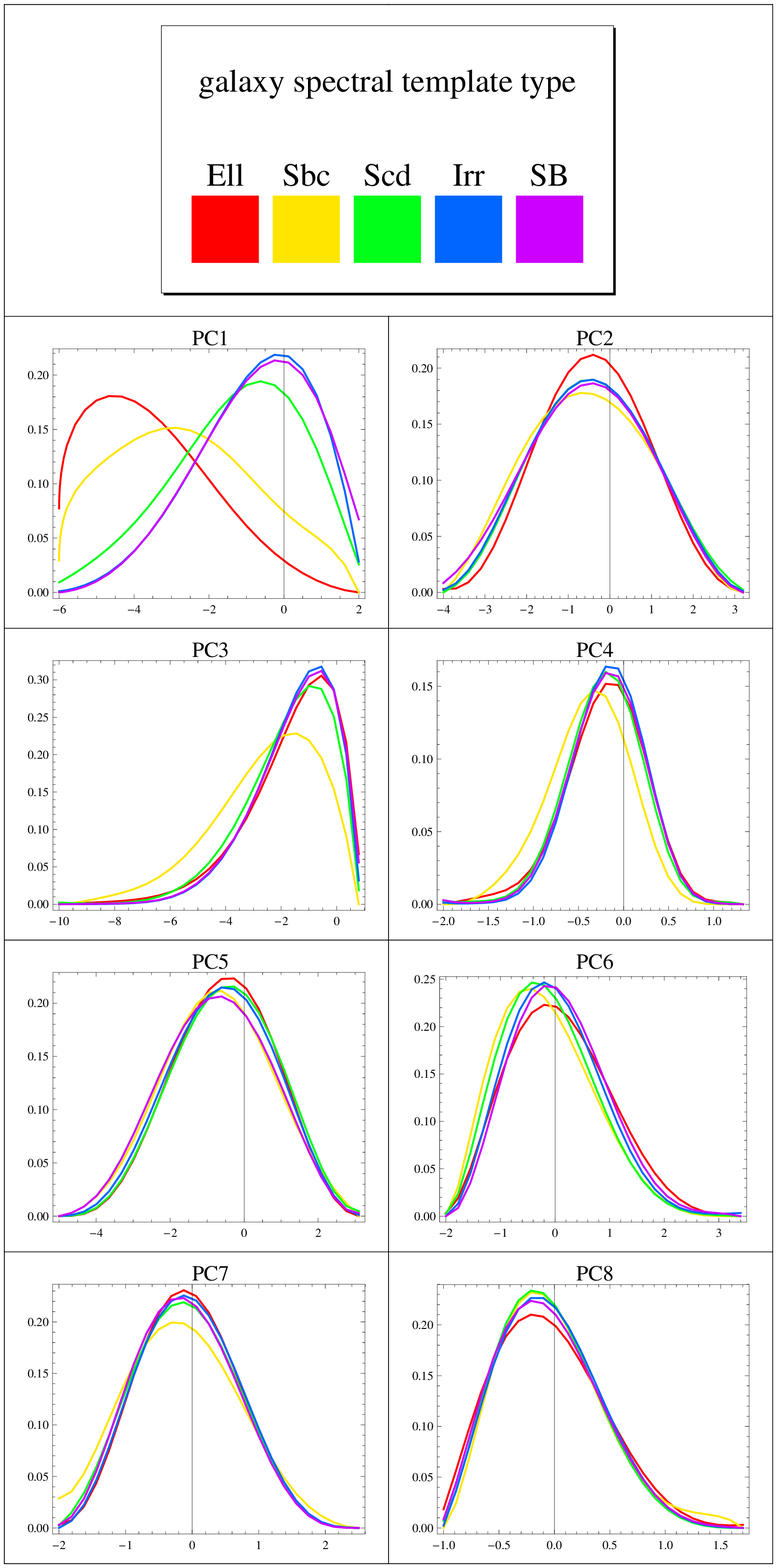}
\caption{Distribution of morphological principal component values for galaxies of different type, for the 2612 galaxies used in this analysis.  Galaxy type is estimated by \citet{lephare}.  }
\label{PCfig}
\end{figure} 

In order to most efficiently determine the effect of the extra information provided by morphological parameters on the photo-z estimation, we form principal components of the morphological parameters.  The morphological principal components are given as linear combinations of the eight morphological parameters discussed in \S \ref{dataset} by Table \ref{tab}.

Principal components \citep[e.g.][]{Jolliffe02} are the result of a coordinate rotation in a multi-dimensional space of possibly correlated data parameters into vectors with maximum orthogonal significance.  The first principal component is along the direction of maximum variation in the data space, the second is along the direction of remaining maximum variation orthogonal to the first, the third is along the direction of remaining maximum variation orthogonal to both of the first two, and so on.

\begin{deluxetable}{lrrrrrrrrr}
\tablecaption{Contents of morphological principal component vectors \label{tab}}
\tabletypesize{\footnotesize}
\startdata

 \nodata & PC1 & PC2 & PC3 & PC4 & PC5 & PC6 & PC7 & PC8 \\
C & -.52 & +.13 & +.01 & -.17 & -.03 & +.19 & .09 & -.8\\
A & +.10 & -.41 & +.62 & -.18 & -.61 & +.15 & +.07 & -.007 \\
S & +.06 & +.29 & +.72 & -.18 & +.60 & -.02 & +.04 & +.03 \\
G & -.47 & -.07 & +.09 & -.19 &  -.09 & -.67 & -.51 & +.12 \\
$M_{20}$ & +.49 & +.05 & +.004 & +.03 & -.09 & -.67 & +.32 & -.44 \\
$\varepsilon$ & +.07 & +.62 & -.14 & -.63 & -.34 & +.03 & +.13 & -.23\\
n & -.50 & -.006 & +.07 & +.21 & -.04 & -.20 & +.74 & +.32 \\
$r_e$ & -.02 & +.58 & +.24 & +.65 & -.36 & +.01 & -.22 & -.03 \\

\enddata
\end{deluxetable}

Given the available galaxy type estimations discussed at the end of \S \ref{dataset}, we can check for correlations between principal components of the morphological parameters and galaxy type, as in Figure \ref{PCfig}.  It is seen that the first principal component is well correlated with galaxy type, and correlations persist through several of the other principal components.  These correlations indicate that the morphology may provide an additional handle on the photo-z estimation, since outliers often occur because a spectral feature (such as a break) of one galaxy type at a given redshift may be seen by the observer to be at the same wavelengths as a spectral feature of another galaxy type at different redshift.  Thus morphological information indicative of galaxy type may help break this degeneracy.

\begin{figure}
\includegraphics[width=3.5in]{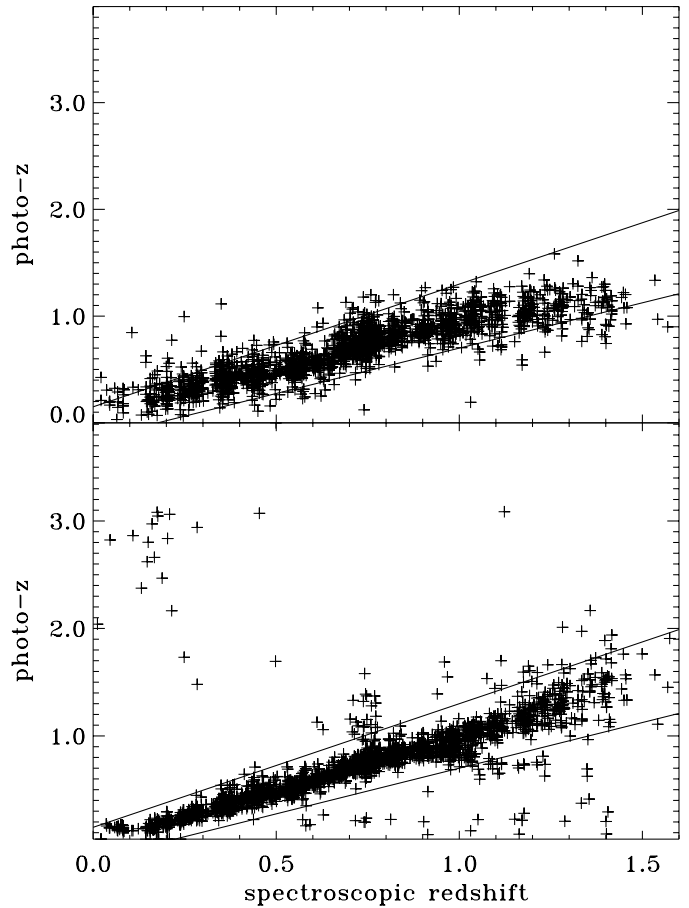}
\caption{{\bf TOP:} The estimated photo-z versus the actual redshift, as determined by the custom artificial neural network used in this work, for the case of no morphological parameters included.  This determination is of the type used in this analysis, with 350,000 training iterations, where the training set is formed from 700 galaxies and the evaluation set, for which the results are plotted, consists of the remaining 1912 galaxies.  `Outliers' in a determination are defined as those where $\vert z_{phot}-z_{spec} \vert \, / \, (1+z_{spec}) > .15$, shown as the two diagonal lines. 
\\{\bf BOTTOM:} The photo-zs for the same galaxies as in the top plot, as estimated with the Lephare template fitting code as reported in \citet{lephare}.  The template fitting method has a lower scatter for non-outliers but a larger number of catastrophic outliers (ie those where $\vert z_{phot}-z_{spec} \vert \, / \, (1+z_{spec}) >> .15$) than the custom neural network for these galaxies.  }
\label{photovsspec}
\end{figure}

\section{Photo-z estimation method} \label{method}

Artificial neural network techniques have been popular empirical methods for photo-z estimation, including with such software packages as ANNz \citep{annz}.  In the case of neural network photo-z determination, the network functions as a `black box' which finds patterns contained in the relation between band magnitudes (and, in principle, other information) and redshift in an unbiased way.  Thus, a neural network photo-z estimation can be a useful tool to explore whether additional parameters beyond the band magnitudes, such as morphology in this case, provide additional useful information.

An artificial neural network, in analogy with a biological one, contains layers of nodes called ``neurons'' and relationships between neurons in different layers of varying weights which can be altered.  Each neuron in the network beyond the input layer assumes a value determined by passing through an activation function the sum of the product of all of the values of the neurons feeding the neuron in question times the weight of the connection between the two neurons.  Neurons at the input accept values of data and the output of the network is the value of one or more output layer neurons.  

The weights between neurons are adjusted by `training' the network to best give desired outputs for a set of inputs.  Training is dependent on a {\it training set} containing a number of cases with inputs and output(s).  In the case of photo-z estimation, the training set contains band magnitudes and possibly other information as inputs, and the actual known (spectroscopic) redshifts as the output.  With the weights set in this way, the network can be used to estimate the redshift of other galaxies in an {\it evaluation set}, and the results can be compared to the known redshifts of the evaluation set to determine the quality of photo-z estimation.  A comprehensive discussion of artificial neural networks is presented in \citet{Haykin99}, and a specialized discussion for the context of photo-z estimation is presented in e.g. \citet{Vanzella04}.

The artificial neural network package used in this analysis is a `multi-layer perceptron' developed for the IDL environment by one of the authors (JS).\footnote{available from www.slac.stanford.edu/$\sim$jacks}  Perceptrons are standard artificial neural network architectures for pattern recognition, consisting of input, hidden, and output neurons as described above.  The primary motivation for the development of this code was to treat additional available galaxy information beyond photometric data (for example shape parameters) on an equal footing with the photometric data.  The IDL code can be relatively easily modified, and could in principle be configured for a wide variety of input data situations.  As training convergence is relatively slow in this network, it is most useful in situations where a robust training set is available from the outset.

As implemented here, the network has an input layer of neurons, five of which accept the observed magnitudes in each optical band, and an additional variable number of input neurons which accept values of as many morphological parameters as desired.  The input layer treats all input information on an equal footing, normalizing each input parameter across all objects in the training set so that the inputs for each neuron on the input layer are distributed between 0 and 1.  There are two hidden layers of 30 neurons each, and an output layer with a single neuron obtaining a value between 0 and 1 which is a proxy for the estimated redshift, with the linear conversion defined during the training when the known redshifts of the training set are supplied subject to the conversion.  

The network uses a hyperbolic tangent activation function for the neurons beyond the input later, and the weights are adjusted during training via the back propagation technique \citep[e.g.][]{Haykin99} where in each training iteration the weights are altered in a way to move `downhill' in the high dimensional surface of summed training set redshift errors in the space of weights.  Each iteration during training consists of the network evaluating the entire training set and adjusting the weights.  In addition to standard back propagation, this network features an algorithm to `kick' the weights away from possible local minima in the summed error.  The top panel of Figure \ref{photovsspec} shows the estimated photo-z versus spectroscopic redshift for the galaxies in the evaluation set of a particular determination with no morphological information included.  This determination features 350,000 training iterations, 700 galaxies in the training set, and 1912 galaxies in the evaluation set, which is the standard used in all determinations here.  The bottom panel of Figure \ref{photovsspec} also shows photo-z estimations for the same galaxies with the Lephare template fitting method as reported in \citet{lephare}.  The custom neural network determination apparently leads to fewer catastrophic outliers (ie those where $\vert z_{phot}-z_{spec} \vert \, / \, (1+z_{spec}) >> .15$) than with the Lephare template fitting method, although has a larger scatter for those galaxies in which the photo-z estimate is close to the actual redshift.

\section{Results on inclusion of shape parameter information} \label{simlumf}

To determine the effect of including a given or multiple principal components in addition to the band magnitudes, we complete six realizations of the training and evaluation process for every case, with a training set of 700 galaxies with 350,000 training iterations and an evaluation set of the remaining 1912 galaxies, and record the number of outliers, and the RMS error, in the evaluation set.  In this work we follow convention \citep[e.g][]{lephare} and define outliers in a given realization as those galaxies where 

\begin{eqnarray}
\nonumber Outliers: {{\vert z_{phot}-z_{spec} \vert} \over {1+z_{spec}}} > .15, 
\label{erroreq}
\end{eqnarray}
where $z_{phot}$ and $z_{spec}$ are the estimated photo-z and actual (spectroscopically determined) redshift of the object respectively.  The RMS photo-z error in a realization is given by a standard definition 

\begin{eqnarray}
\nonumber \sigma_{\Delta z/(1+z)} \equiv \sqrt { {{1} \over {n_{gals}}}  \Sigma_{gals} \left( {{ z_{phot}-z_{spec} } \over {1+z_{spec}}} \right) ^2 }, 
\label{RMSeq}
\end{eqnarray}
where $n_{gals}$ is the number of galaxies in the evaluation set and $\Sigma_{gals}$ represents a sum over those galaxies.  Note that we do not exclude outliers from the calculation of the RMS photo-z error.  Because in each realization the membership of the training set varies, and because as the training process contains `kicks' to knock the weights away from local minima in the summed error (see \S \ref{method}), each realization for a given input parameter set produces a slightly different number of galaxies in the evaluation set with outlier level errors, and a slightly different average error.  For comparison, the template fitting results reported by \citet{lephare} give 5\% outliers and an RMS error of $\sigma_{\Delta z/(1+z)} =.1881$ for this sample.  This error is dominated by the catastrophic outliers (Figure \ref{photovsspec}), and drops substantially to below that of the custom neural network method if outliers are excluded.

\begin{figure}
\includegraphics[width=3.5in]{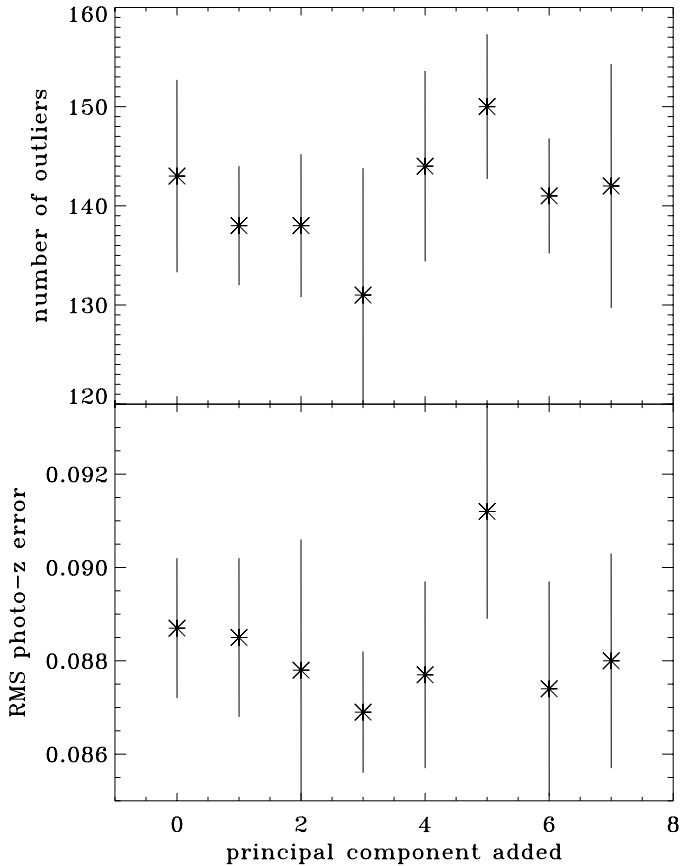}
\caption{Number of outliers (TOP) and RMS error $\sigma_{\Delta z/(1+z)}$ (BOTTOM) in the photo-z estimation with the inclusion of different individual principal components of the morphological parameters.  The uncertainties represent the standard deviation of the values obtained from different realizations, as discussed in \S \ref{simlumf}.  }
\label{indpcs}
\end{figure} 

\begin{figure}
\includegraphics[width=3.5in]{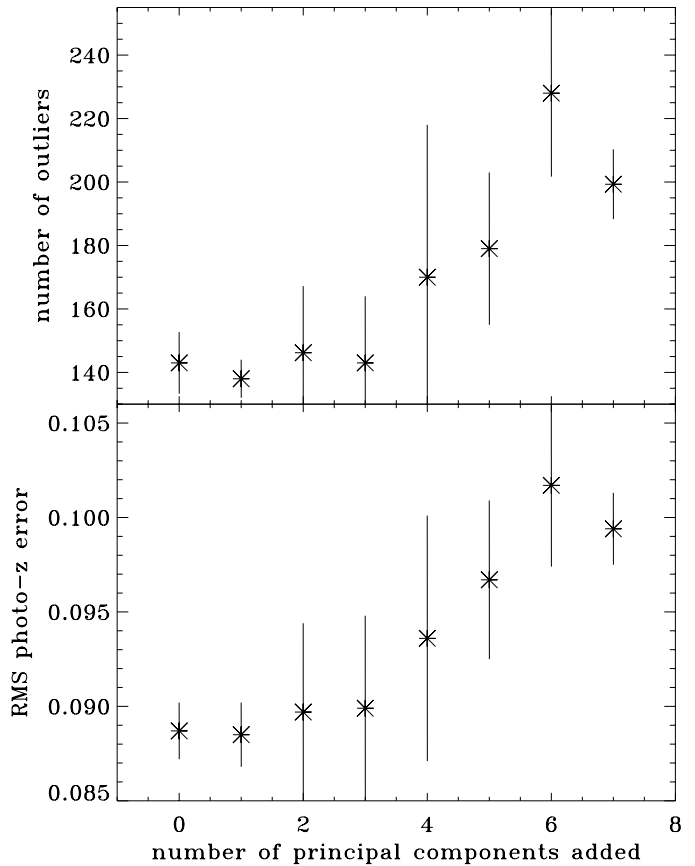}
\caption{Number of outliers (TOP) and RMS error $\sigma_{\Delta z/(1+z)}$ (BOTTOM) in the photo-z estimation  with the inclusion of multiple principal components of the morphological parameters, starting with none, adding the first morphological principal component, then the first and the second principal components, and so on.  The uncertainties represent the standard deviation of the values obtained from different realizations, as discussed in \S \ref{simlumf}. }
\label{multpcs}
\end{figure} 

Figure \ref{indpcs} shows the number of outliers and RMS error for the inclusion of the seven different principal components individually.  Figure \ref{multpcs} shows the number of outliers and RMS error for the inclusion of multiple principal components, starting with none, then adding in the first, then adding in the first and second, then adding in the first through third, and so on.  In each figure, the error bars correspond to the standard deviation of the number of outliers or RMS scatter in the different realizations.  We note that with six realizations per case, the standard deviations in the number of outliers and RMS photo-z error are not particularly robust, however we include them to provide a sense of the scatter of results from different realizations.  We note that the last principal component (PC8) should by definition contain minimal significant variation in the morphological parameters, so we do not include it in the analysis.

It is apparent that adding in any one of the principal components of the morphological parameters may provide a small decrease in the average number of outliers or the RMS error or both, but the differences are not statistically significant compared to the inclusion of no morphological information.  As seen in Figure \ref{multpcs}, adding multiple principal components increases the number of outliers and the RMS error.

\section{Discussion}

We have used a custom artificial neural network for photometric redshift estimation to evaluate the effects of including galaxy shape information, in the form of principal components of morphological parameters, on the photo-z estimation.  The data set we use consists of 2612 galaxies with five optical band magnitudes, reliable spectroscopic redshifts, and eight morphological parameters each.  We note that a neural network is in a sense an unbiased way of determining the relative strength of the correlations of a set of input parameters with the output parameter, and that this network is designed to treat all input parameters on an equal footing.

In order to more effectively include the morphological information, we form principal components of the morphological parameters.  An analysis of the principal components and galaxy types shows that the value of the first few principal components and galaxy type are correlated.  

However, we find that the inclusion of morphological information does not significantly decrease the number of outliers or RMS error in photo-z estimation for this data set with the neural network technique used.  When only one principal component of the morphological parameters is included, there can be a slight but not significant decrease.  When multiple principal components of the morphological parameters are included, the number of outliers and the RMS error increases.  We conclude that any gain that may arise in a neural network photo-z determination from correlations between morphology and redshift in this data set is overwhelmed by the additional noise introduced.  It may be that any correlations between principal components of the morphological parameters and the galaxy type are degenerate to some extent with the correlations between galaxy type and galaxy colors.

This analysis is applicable to artificial neural network photo-z estimations, and possibly other training set methods, with similar data.  It is possible, however, that morphological parameters could yield improvements for other algorithms, especially template-fitting methods with relatively large outlier fractions.  This is because such outliers usually occur when a particular spectral break is confused for another break in a different galaxy type at a different redshift (e.g. an elliptical galaxy at low redshift mistaken for a spiral at high redshift).  Having additional information to guide the template selection might therefore be helpful in reducing the outlier fraction.  A preliminary analysis using Figure \ref{PCfig} to build prior probability distributions on the template selection in the Lephare package produced a few percent reduction in the number of outliers.  A more thorough analysis of the effects of shape parameters in photometric redshift estimation with a template fitting method will be presented in a forthcoming work.   

\acknowledgments

JS thanks T. Brookings for his counsel, and S. Kahn and R. Schindler for their encouragement and support.  MS is thankful to R. Blandford and P. Marshall for very useful discussions and support.  This work was supported in part by the U.S. Department of Energy under contract number DE-AC02-76SF00515.


\begin{thebibliography}{}

\bibitem[Arnouts et al.(1999)]{Arnouts99} Arnouts, S., Cristiani, S., Moscardini, L., Matarrese, S., Lucchin, F., Fontana, A., \& Giallongo, E. 1999, \mnras, 310, 540
\bibitem[Ben{\'{\i}}tez(2000)]{bpz} Ben{\'{\i}}tez, N.\ 2000, \apj, 536, 571 
\bibitem[Collister \& Lahav(2004)]{annz} Collister, A. \& Lahav, O. 2004, \pasp, 116, 345
\bibitem[Davis et al.(2007)]{EGS} Davis, M., et al. 2007, \apj, 660, L1
\bibitem[Gerdes et al.(2010)]{Gerdes10} Gerdes, D., et al. 2010, \apj,  715, 823
\bibitem[Graham \& Driver(2005)]{GD05} Graham, A. \& Driver, S. 2005, PASA, 22, 118
\bibitem[Griffith et al.(2011)]{Griffith11} Griffith, R., et al. 2011, in prep
\bibitem[Gwyn (2008)]{Gwyn08} Gwyn, S., 2008, \pasp, 120, 212
\bibitem[H\"ausler et al.(2007)]{Hausler07} H\"ausler, B., et al. 2007, \apjs, 172, 615
\bibitem[Hearin et al.(2010)]{Hearin10} Hearin, A., Zentner, A., Ma, Z., \& Huterer, D. 2010, \apj, submitted, (arXiv:1102.3383)
\bibitem[Haykin(1999)]{Haykin99} Haykin, S., {\it `Neural Networks: A Comprehensive Foundation'}, Upper Saddle River, NJ: Prentice Hall 1999
\bibitem[Huterer et al.(2006)]{Huterer06} Huterer, D., Takada, M., Bernstein, G., \& Jain, B. 2006, \mnras, 366, 101
\bibitem[Ilbert et al.(2006)]{lephare} Ilbert, O., et al. 2006, \aap, 457, 841
\bibitem[Ivezic et al.(2008)]{LSSTover} Ivezic, Z. et al. 2008 arXiv:0805.2366
\bibitem[Jolliffe(2002)]{Jolliffe02} Jolliffe, I., {\it `Principal Component Analysis'}, Series: Springer Series in Statistics, 2nd ed. New York, NY: Springer 2002
\bibitem[Koekemoer et al.(2007)]{K07} Koekemoer, A., et al. 2007, \apjs, 172, 196
\bibitem[Lotz et al.(2008)]{Lotz08} Lotz, J., et al. 2008, \apj, 672, 177
\bibitem[Peng et al.(2002)]{Peng02} Peng, C., Ho, L., Impev, C., \& Rix, H. 2002, \aj, 124, 266
\bibitem[Scarlata et al.(2007)]{Scar06} Scarlata, C., et al. 2007, \apjs, 172, 406
\bibitem[Tagliaferri et al.(2003)]{Tag03} Tagliaferri, R., et al. 2003, {\it Lect. Notes Comp. Sci.}, 2859, 226
\bibitem[Vanzella et al.(2004)]{Vanzella04} Vanzella, E., et al. 2004, \aap, 423, 761
\bibitem[Vince \& Csabai(2006)]{VC2006} Vince, O. \& Csabai, I. 2006, `Toward more precise photometric redshift estimation.' Proceedings of the International Astronomical Union, 2, pp 573-574 
\bibitem[Way \& Srivastava(2006)]{WS2006} Way, M. \& Srivastava,  A. 2006, \apj, 647, 102

\end{thebibliography}
\end{document}